\documentclass[twocolumn,aps,showpacs,amsmath,amssymb]{revtex4}

\usepackage{graphicx}% Include figure files
\usepackage{bm}% bold math
\begin{document}

\title{Chaotic stirring by a mesoscale surface-ocean flow}

\author{Edward R. Abraham} \email{e.abraham@niwa.cri.nz}
\affiliation{
  National Institute of Water and Atmospheric Research (NIWA), P.O.
  Box 14-901, Kilbirnie, Wellington, New Zealand
}
\homepage{http://www.niwa.cri.nz} \author{Melissa M. Bowen}
\email{m.bowen@niwa.cri.nz} \affiliation{University of Colorado,
  Boulder, Colorado, 80309-0429}
\altaffiliation[Present address:]{
  National Institute of Water and Atmospheric Research (NIWA), P.O.
  Box 14-901, Kilbirnie, Wellington, New Zealand
}
\date{June 10, 2002. \emph{Chaos} {\bf 12} 373 - 381 (2002) doi:10.1063/1.1481615}

\begin{abstract}
  The horizontal stirring properties of the flow in a region of the
  East-Australian Current are calculated. A surface velocity field
  derived from remotely sensed data, using the Maximum Cross
  Correlation method, is integrated to derive the distribution of the
  finite-time Lyapunov exponents. For the region studied (between
  latitudes 36$^\circ$S and 41$^\circ$S and longitudes 150$^\circ$E
  and 156$^\circ$E) the mean Lyapunov exponent during 1997 is
  estimated to be $\lambda_\infty = 4 \times 10^{-7}$ s$^{-1}$. This
  is in close agreement with the few other measurements of stirring
  rates in the surface ocean which are available. Recent theoretical
  results on the multifractal spectra of advected reactive tracers are
  applied to an analysis of a sea-surface temperature image of the
  study region. The spatial pattern seen in the image compares well
  with the pattern seen in an advected tracer with a
  first-order response to changes in surface forcing. The response timescale is estimated to be 20 days.
\end{abstract}

\maketitle

{\bf\noindent Satellite imagery of the ocean surface shows that sea-surface
temperature and chlorophyll have complex distributions. The filaments
and whorls characteristic of tracers in stirred fluids are often
clearly evident. Despite the importance of stirring to a range of
problems, such as plankton ecology, larval transport or predicting
the fate of pollutants, it has not been well described. Here we
analyse the stirring properties of a velocity dataset obtained from
the interpretation of sequential satellite images. The finite-time
Lyapunov exponents of the flow are determined. Using these results, we
show that the patterns seen in a sea-surface temperature image are
consistent with those formed from the advection of a reactive
tracer. This suggests that an understanding of the distribution of
tracers such as temperature and chlorophyll in the surface ocean must
include a representation of both the chaotic advection by the flow and
the dynamics of the tracers themselves. If the advection can be characterised, as we have been able to do here, then it may be possible to use the
information contained in sea-surface imagery to infer the
time-scales of the tracer dynamics.}

\section{\label{sec:introduction}Introduction}

 One of the simplest measures of the distribution of a
tracer is its power-spectrum, which quantifies the variability in the
distribution at each spatial scale. Between inverse wavenumbers of $1$
and $100$ km the spectra of sea-surface temperature and chlorophyll are
usually found to have a power-law form, $F(k) \sim k^{-\beta}$. The
exponent, $\beta$, is typically in the range $1.5< \beta < 2.5$
\cite{Gower80, Deschamps81, Denman88, Ostrovskii95, Piontovski97,
  Washburn98}.  Where both sea-surface temperature and chlorophyll are
measured at the same time their spectra often have similar slopes.
Ocean currents are turbulent within the mesoscale wavenumber range, 1
km $< k^{-1} <$ 100 km, and their energy falls off rapidly at higher
wavenumber \cite{Monin85}. The flow within this range is approximately
two dimensional, and it is appropriate to consider the advection of a
passive tracer in the surface ocean as a stirring process
\cite{Garrett83, Pierrehumbert94}. Stirring acts to stretch and fold
patches of tracer, distorting them into long tendrils and filaments.
In spectral terms, variance that is input at low wavenumber is
transferred by stirring towards higher wave-numbers, where it is
dissipated by diffusion. The spectrum of a stirred tracer, which has a
source at low wavenumber but is otherwise conserved, is expected to be
power-law with an exponent $\beta = 1$ \cite{Batchelor59}. Batchelor
law spectra with this form are sometimes seen in laboratory
experiments of stirring by two-dimensional flows \cite{Wu95}, but
$\beta=1$ is outside the range usually measured for sea-surface
temperature and chlorophyll, as has been noted by several authors
\cite{Gower80, Deschamps81}. It is clear, however, that neither
sea-surface temperature or chlorophyll are conserved quantities.  For
a non-conserved tracer which satisfies a first-order equation, any
spectral exponent $\beta >1$ can be obtained, depending on the
relation between the reaction-rate and the typical stirring-rate of
the flow \cite{Corrsin61}. This suggests that the patterns seen in the
surface ocean may reflect both the stirring by the flow and the
tracers' non-conservation. Recently, results have been derived which
give the multifractal structure of an advected reactive tracer in
terms of the distribution of finite-time Lyapunov exponents, which
characterise the stirring by the flow \cite{Neufeld00}.  The
theoretical results were derived in the context of Lagrangian chaotic
flows. Applying the theory to the surface ocean is hampered by the
lack of data on stirring by ocean currents. The few measurements which
have been made depend on following the stretching of a tracer patch
for an extended period of time \cite{Ledwell93, Ledwell98, Abraham00}.
This technique gives little information on the distribution of
stirring rates, each observation of a tracer patch typically returning
only a single estimate of the stirring. Given the lack of direct
measurents, an indirect approach must be taken. In this paper, surface
velocity data, inferred from satellite observations, are used to
calculate stirring rates. The analysis is focussed on the East
Australian region, where there is strong mesoscale eddy activity and
where a suitable dataset is available. The derived distribution of
stirring rates, more formally finite-time Lyapunov exponents, is
applied to an analysis of the distribution of sea-surface temperature.

\section{\label{sec:data} Surface velocity data}

\begin{figure}
  \includegraphics{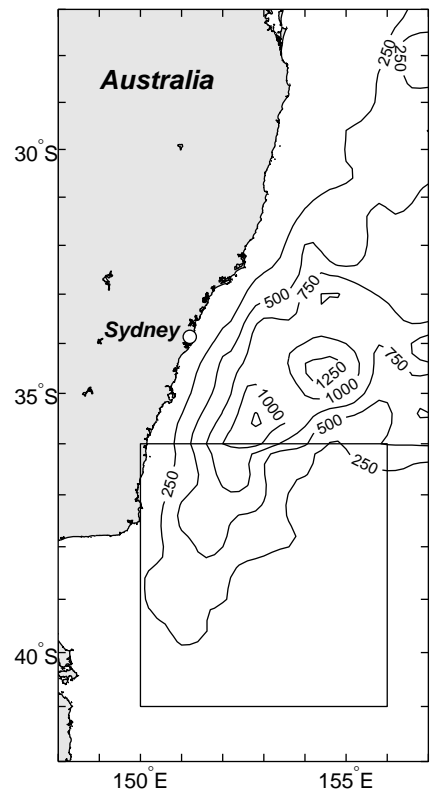}
\caption{\label{fig:location} The study region in the south-west
  Tasman Sea, showing the region over which the satellite derived
  velocity data is available. The contours show the average eddy
  kinetic energy (half the velocity variance, cm$^2$ s$^{-2}$) from
  the 1997 data. The small box marks the area over which stirring
  rates are derived.}
\end{figure}

It is possible to estimate sea-surface velocity from a comparison
between sequential satellite images of sea-surface temperature
\cite{Emery86, Schmetz87, Kelly92}. This Maximum Cross-Correlation
(MCC) method has recently been applied to 7 years of satellite data
from 1993 to 1999 \cite{Bowen01}, covering the region where the East
Australian Current (EAC) separates from the coast and heads across the
Tasman Sea. The MCC data has a high spatial and temporal resolution,
but because no sea-surface temperature imagery can be obtained through
cloud, there are extensive data dropouts. This problem has previously
limited the application of the MCC method to sequences of a few
exceptional cloud free images. In the dataset used here, the MCC
velocities have been blended with lower resolution data derived from
an analysis of satellite-altimeter measured sea-surface height
\cite{Wilkin01}. The resulting optimally interpolated (OI) surface
velocity field has a uniform spatial and temporal resolution of 100 km
and 10 days, sufficient to resolve mesoscale features. While both the
satellite altimeter and temperature data are globally available, they
have only been combined to derive surface velocities in the region
that is studied here. In this paper the stirring properties of the OI
flow are characterised.

The EAC current is an intense western boundary current which flows
south, close to the Australian coastline, until it reaches the
latitude of Sydney. Here the current turns, forming large loops that
pinch off to form eddies. The eddies continue moving south while the
main current turns to flow east. The separation region is seen as a
peak in the eddy kinetic energy of the flow (Fig.~\ref{fig:location}).
Analysis is restricted to a smaller region to the south of the main
separation zone, between latitudes 36$^\circ$S and 41$^\circ$S, and
between longitudes 150$^\circ$E and 156$^\circ$E. The mean currents in
this region are relatively weak, so trajectories remain in the area
where the OI velocities are defined for long enough to allow the
stirring to be determined.

In the following two sections, results on stirring in two-dimensional
flows are briefly reviewed.

\section{\label{sec:lyapunov}Finite-time Lyapunov exponents}

In a two-dimensional divergence-free flow, with velocity
$\bm{v}=(u,v)$, the change in the separation between the trajectories
of two infinitesimally separated points $\bm{x}(t)$ and $\bm{x}(t) +
\delta\bm{x}(t)$ is
\begin{equation}
\label{eq:dx_dt}
\dot{\delta\bm{x}}  = \bm{S}(t)\delta \bm{x},
\end{equation} 
where the matrix $\bm{S}(t)$ is the Jacobian,
\begin{equation}
\bm{S}(t) = 
\left(
\begin{array}{cc}
\partial_x u  & \partial_y u \\
\partial_x v  & \partial_y v \\
\end{array}
\right),
\end{equation}
and is evaluated along the trajectory ${\bm x}(t)$ \cite{Ottino89}.
The Jacobian may be resolved into a symmetric and antisymmetric part.
The antisymmetric part is a pure rotation. The symmetric part is the
strain tensor,
\begin{equation}
\bm{E}(t)=\left(
\begin{array}{cc}
\partial_x u & (\partial_y u + \partial_x v)/2 \\
(\partial_y u + \partial_x v)/2 & -\partial_x u \\
\end{array}
\right),
\end{equation} which acts to stretch a small patch of tracer without
changing its area. The strain tensor has the eigenvalues $\gamma_{\pm}
= \pm \sqrt{\partial_x u^2 + (\partial_y u + \partial_x v )^2/4}$,
with the corresponding eigenvectors $\bm{\gamma}_\pm$. In a
pure-strain flow with velocity $\bm{v} = (\gamma x,-\gamma y)$ the
strain tensor is diagonal, and the solution to Eq.~(\ref{eq:dx_dt}) is
$\delta \bm{x} = (\delta x(0) e^{\gamma t}, \delta y(0) e^{-\gamma
  t})$. There is an exponential growth in the component of the
separation which is aligned with the positive eigenvector, and an
exponential decay of the other component. In this flow field a small
initially circular patch of tracer is stretched into an ellipse, with
the rate of growth of the major axis being given by the strain-rate,
$\gamma$.

In a general time-dependent flow infinitesimal separations between
particles will still grow exponentially with time, but the growth is
no longer a simple function of the strain \cite{Ott93}. It may be
characterised by the finite-time Lyapunov exponents, defined by
\begin{equation}
  \label{eq:lambda_t}
  \lambda(\bm{x}(t), t) = {1 \over t}\log\left( {|\delta \bm{x}(t)|
  \over |\delta \bm{x}(0)|}\right),
\end{equation}
where the orientation of the initial separation, $\delta \bm{x}(0)$, is chosen so that
$\lambda$ is maximal. At large times, $\lambda$ may be approximated by
using Eqs.~(\ref{eq:dx_dt}) and (\ref{eq:lambda_t}) with a randomly chosen
initial orientation of $\delta \bm{x}$, as the stretching by the flow
aligns most initial vectors with the direction of maximal stretching.
We refer to the Lyapunov exponents approximated in this way as
$\tilde\lambda$. To calculate the Lyapunov exponents at small times
more care must be taken.  An infinitesimal circle is deformed by the
flow into an ellipse. The integrated deformation along a trajectory,
$\bm{M}(t)$, may be obtained by solving
\begin{equation}
\label{eq:mdot}
\dot{\bm M} = {\bm S}{\bm M},
\end{equation} 
where ${\bm M}(0)$ is the identity matrix. The semi-major and
semi-minor axes of the ellipse are the eigenvectors of the matrix
${\bm M}^{\text{T}}{\bm M}$, with their squared lengths being the
corresponding eigenvalues (denoted $m_+$ and $m_-$, with $m_+$
referring to the larger value). The finite-time Lyapunov exponent
along a trajectory is then given by
\begin{equation}
  \label{eq:lambda_m}
  \lambda(\bm{x}(t), t) = {1 \over 2t}\log(m_+),
\end{equation}

In a closed ergodic flow the Lyapunov exponents converge with time
towards a single value, $\lim_{ t \to \infty} \lambda(t) =
\lambda_\infty$. If the flow is chaotic then $\lambda_\infty > 0$. At
times $t>>1/\lambda_\infty$ the averaged finite-time Lyapunov exponent
has the form
\begin{equation}
  \label{eq:lambda_convergence}
  \langle\lambda\rangle \sim A/t + B/\sqrt{t} + \lambda_\infty, 
\end{equation}
where $A$ and $B$ are two constants \cite{Tang96}. This relation has
previously been used to estimate $\lambda_\infty$ where the
integration time was too short to allow convergence of the finite-time
exponents to be achieved \cite{Thiffeault01}. The long-time behaviour
of the standard deviation of the finite time Lyapunov exponents,
$\sigma_\lambda$, is given by
\begin{equation}
  \label{eq:sigma_convergence}
  \sigma_\lambda \sim \sqrt{\Delta/t}, 
\end{equation}
where $\Delta$ is a third constant \cite{Tang96}. The probability
distribution of finite-time Lyapunov exponents has the time-asymptotic
form
\begin{equation}
\label{eq:p_lambda}
P(\lambda(t), t) \sim t^{1/2}e^{-G(\lambda)t},
\end{equation}
where $G(\lambda) > 0$ and $G(\lambda_\infty)=G'(\lambda_\infty)=0$
\cite{Ott93, Bohr98}. If $P(\lambda, t)$ is gaussian then $G(\lambda)$
is a parabola,
\begin{equation}
\label{eq:g_parabola}
G(\lambda) = {(\lambda-\lambda_\infty)^2 \over 2\Delta}.
\end{equation}

\section{\label{sec:scalar_spectra}The distribution of advected tracers}

As a simple approximation, sea-surface temperature may be regarded as
dynamically passive, with horizontal temperature gradients not
affecting the flow. This view is supported by observations which show
that, away from major fronts, horizontal temperature variations within
the surface layer are matched by salinity variations in such a way
that there are only weak horizontal density gradients \cite{Chen95,
  Rudnick99, Ferrari01}. As a parcel of surface water is advected by
the flow it exchanges heat with the atmosphere and with the underlying
water. An initial exploration of the dynamics of sea-surface
temperature may be made by assuming that the changes in temperature
are first-order.  Consider a tracer $C(\bm{x}(t),t)$ which satisfies
the Lagrangian equation
\begin{equation}
\label{eq:dcdt}
{dC \over dt} = \alpha(C_0 - C),
\end{equation}
where $\alpha$ is a relaxation rate. The source, $C_0(\bm{x},t)$, is
assumed to only vary over large scales, i.e.~there is a wavenumber
$k_0$ such that the fourier power spectrum, $|\hat C_0(k)|^2$, is zero
for $k>k_0$.  If the tracer is advected by a flow which has
$\lambda(t)=\lambda_\infty$ and which has a horizontal diffusivity
$D$, then for wavenumbers between $k_0$ and the diffusive cut-off,
$k_D = \sqrt{\lambda_\infty/D}$, the power spectrum has a power-law
form \cite{Corrsin61},
\begin{equation}
 |\hat C(k)^2| \sim k^{-\beta}.
\end{equation}
The power-law exponent is
\begin{equation}
\label{eq:beta_monofractal}
 \beta=1 + {2\alpha \over \lambda_\infty}.
\end{equation}
For tracers with a rapid relaxation time there is less transfer of
variance toward higher wavenumber and the power-spectrum is steep,
whereas for tracers with a slow relaxation rate the spectrum
approaches that of a forced conserved tracer with Batchelor law
scaling, $\beta = 1$ \cite{Batchelor59}.

For more general flows with a distribution of finite-time Lyapunov
exponents (Eq.~\ref{eq:p_lambda}) the tracer distribution is
multifractal, with the scaling exponents being related to the function
$G(\lambda)$.  Multifractal analysis has previously been carried out
on small-scale temperature and phytoplankton data
(e.g.~\cite{Seuront96, Lovejoy01}). Here theoretical results obtained
by Neufeld {\it et al.~} \cite{Neufeld00} are presented. The $q$th order
structure function is defined as
\begin{equation}
\label{eq:Sq}
S_q(\delta r)= \left\langle\left|C(\bm{x}+\delta\bm{x},t) -
    C(\bm{x},t)\right|^q\right\rangle,
\end{equation}
where $\delta r = |\delta \bm{x}|$ and the angle brackets indicate
averaging over spatial locations and separations $\delta{\bm x}$. For
a scale-free distribution, the structure functions have a power-law
form in the limit $\delta r \rightarrow 0$,

\begin{equation}
S_q(\delta r)\sim \delta r^{\zeta_q}.
\end{equation}
For an advected first-order tracer (Eq.~\ref{eq:dcdt}) the scaling
exponents are given by
\begin{equation}
\label{eq:zetaq}
\zeta_q = \min\left\{q,{q\alpha+G(\lambda)\over\lambda}\right\},
\end{equation}
where the minimum is taken over all values of $\lambda>0$. The
power-spectral exponent is related to the structure functions by the
relation
\begin{equation}
\label{eq:beta_zeta}
\beta = \zeta_2 +1.
\end{equation}In cases
where the function $G(\lambda)$ may be approximated by a parabola (Eq.
\ref{eq:g_parabola}) the scaling exponents have the form
\begin{equation}
\label{eq:zetaq_ld}
\zeta_q=\min\left\{q,  {\lambda_\infty \over \Delta}(\sqrt{1 + 2q\alpha\Delta/\lambda_\infty^2}-1)\right\}.
\end{equation}
In the limit $\Delta \rightarrow 0$, $P(\lambda)$ becomes a delta
function, $\delta(\lambda - \lambda_\infty)$, and the tracer
distribution is monofractal with the scaling exponents
\begin{equation}
\label{eq:zetaq_mono}
\zeta_q =
q\alpha/\lambda_\infty ({\text{for }} \alpha<\lambda_\infty).
\end{equation} 
When $q=2$ the equation for the power-spectral exponent,
Eq.~(\ref{eq:beta_monofractal}), is recovered using
Eq.~(\ref{eq:beta_zeta}). The scaling exponents have the same form
(Eq.~\ref{eq:zetaq_mono}) in the $q\rightarrow 0$ limit. This may be
used to estimate $\alpha/\lambda_\infty$ even if the function
$G(\lambda)$ is unknown.

\section{\label{sec:lyapunov_OI}Finite-time Lyapunov exponents of the OI data}

\begin{figure*}
  \includegraphics{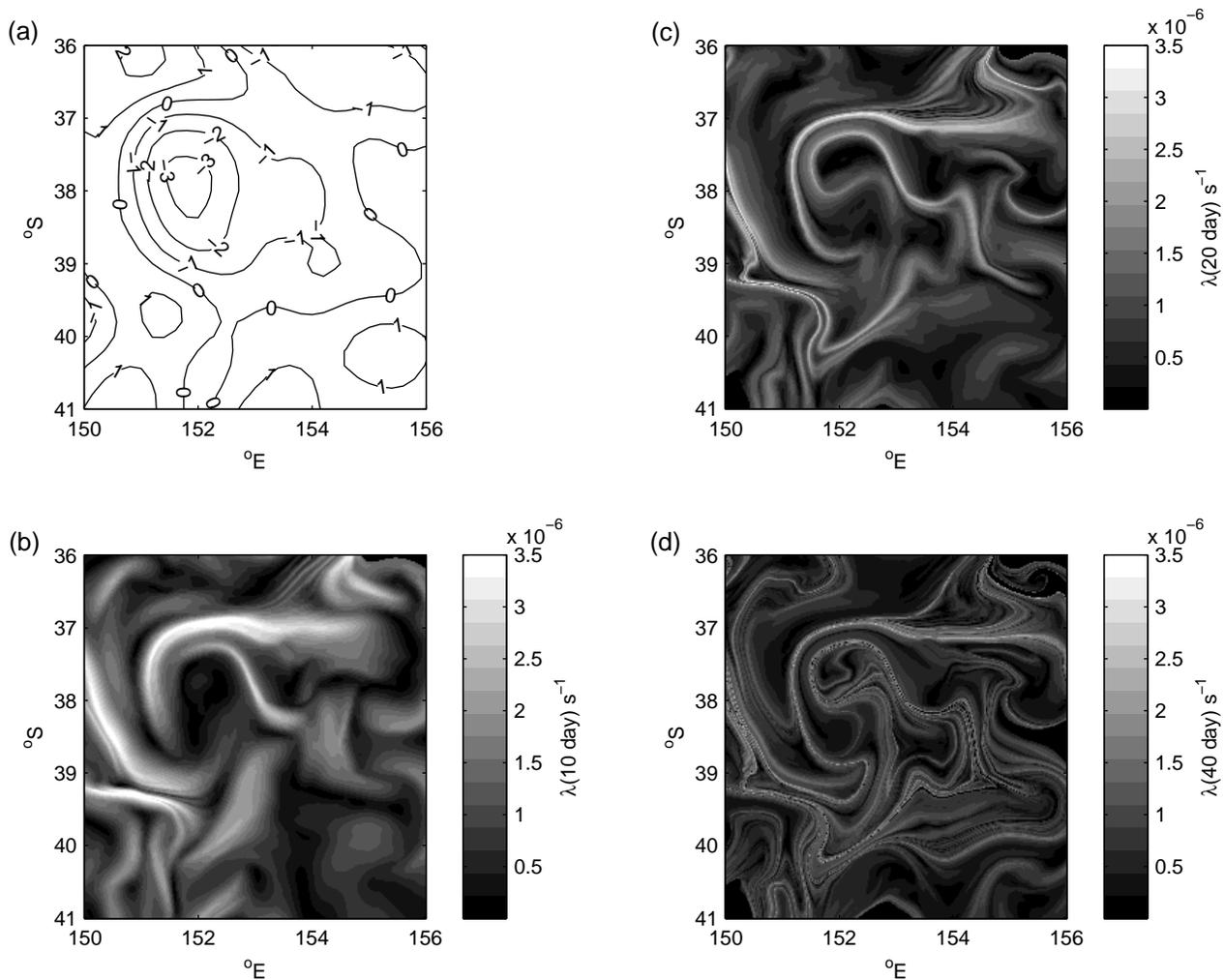}
\caption{\label{fig:lyapunov_hires} The streamfunction and the
  finite-time Lyapunov exponents within the study region on 22
  November 1997. (a) the streamfunction, $\psi$ ($\times 10^4$ m$^2$
  s$^{-1}$) (b) $\lambda(t)$, $t=10$ days (c) $\lambda(t)$, $t=20$
  days (d) $\lambda(t)$, $t=40$ days. }
\end{figure*} 

\begin{figure}
  \includegraphics{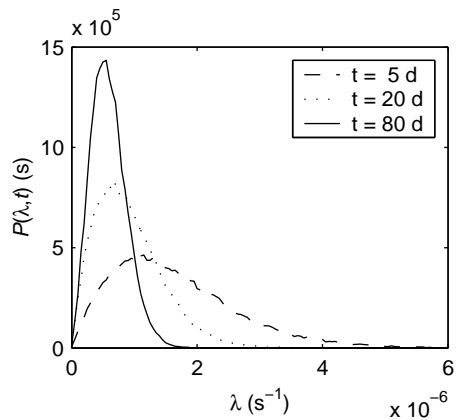}
\caption{\label{fig:p_lambda} The probability distribution of the finite-time
  Lyapunov exponents. As expected the distribution narrows with time,
  but even after 80 days the distribution has not converged.}
\end{figure} 

\begin{figure}
  \includegraphics{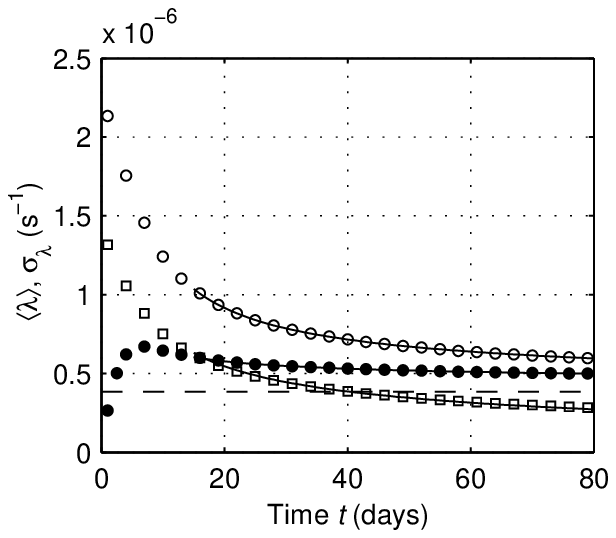}
\caption{\label{fig:lambda_mean_std} The variation with time of the
  mean (circles) and standard deviation (squares) of the distribution
  of finite-time Lyapunov exponents. The open symbols are from the
  distribution of the maximum Lyapunov exponents, $\lambda$. The solid
  symbols are from the approximation to the Lyapunov exponents,
  $\tilde\lambda$, calculated from the stretching of randomly aligned
  initial vectors using Eq.~(\ref{eq:lambda_t}). The solid lines show
  the least-squares fit to the data between 15 and 80 days, using
  Eqs.~(\ref{eq:lambda_convergence}) and (\ref{eq:sigma_convergence}).
  The horizontal dashed line marks the asymptotic value of the
  Lyapunov exponent, \hbox{$\lambda_\infty$ = 3.84 $\times$ 10$^{-7}$
    s$^{-1}$}. }
\end{figure} 

The finite-time Lyapunov exponents are calculated for the study region
using the 1997 OI velocity data. A 4th order Runge-Kutta integration
with a daily time-step is used to calculate the trajectories of a
regular grid of points. The OI velocities are objectively mapped onto
a regular grid with a 20 km by 20 km by 5 day spacing, and the data
are linearly interpolated in time and space to estimate the velocity
along the trajectories. The Jacobian is integrated following
Eq.~(\ref{eq:mdot}) to obtain the matrix ${\bm M}^{\text{T}}{\bm M}$,
and the Lyapunov exponents are then calculated from the eigenvalues
$m_+$ (Eq.~\ref{eq:lambda_m}). There are two components to the OI
velocity field, the mean flow and the velocity anomaly. The optimal
interpolation ensures that the anomaly is divergence free, but the
mean flow is derived directly from the MCC data and does not have this
constraint. While the full velocity field is used to derive the
trajectories, only the traceless component of the Jacobian is
integrated to obtain $\bm{M}$. As an example of the results, the flow
is shown on the 22 November 1997 in Fig.~\ref{fig:lyapunov_hires},
along with the Lyapunov exponents for three different times $t$. In
this figure a regular grid of points, with a spacing of 2.2 km, was
integrated backwards in time.  Trajectories that left the region
within which the OI velocities were defined are shown in black. The
Jacobian was then integrated forwards along the trajectories to obtain
the Lyapunov exponents at a regular grid of points. The streamfunction
on November 22 is dominated by an anti-cyclonic eddy with a diameter
of approximately 200 km, centred on 38$^\circ$S 152$^\circ$E. The ten
day Lyapunov exponents show that the eddy centre is an area of
relatively low stretching, with the highest stretching rates being in
arms around the eddy. For larger times $t$ the Lyapunov exponent
develops a filamentary structure, with the width of the filaments
narrowing with time.

The probabability distribution $P(\lambda,t)$ is shown in
Fig.~\ref{fig:p_lambda} for various times $t$. This was calculated
from sets of forward trajectories initialised on a 61 $\times$ 51
regular grid. Each set of trajectories began 10 days apart, between 1
January and 12 October 1997. As expected, the distribution of the
exponents narrows with time as each trajectory experiences a range of
flow conditions. The integration is not continued beyond 80 days as by
that time over a third of the trajectories have left the region where
the OI velocities are available. The strain of the OI flow in the
study region has the mean and standard deviation
$\langle\gamma_+\rangle = 2.5 \times 10^{-6}$ s$^{-1}$ and
$\sigma_{\gamma_+} = 1.6 \times 10^{-6}$ s$^{-1}$. The initial
stretching is at the rate given by the strain, but the average
Lyapunov exponent decreases with time
(Fig.~\ref{fig:lambda_mean_std}).  The mean Lyapunov exponent is still
decreasing at the end of the 80 day integration so the form of
$G(\lambda)$ cannot be determined. The Lyapunov exponent
$\lambda_\infty$ can only be estimated by fitting the relation for the
expected time-dependence, Eq.~(\ref{eq:lambda_convergence}) to the
data.  The least-squares fitted curve to the mean Lyapunov exponent
between 15 and 80 days is $\langle\lambda\rangle \sim 0.371 / t + 4.17
\times 10^{-4}/\sqrt{t} + 3.84 \times 10^{-7}$, where the time, $t$,
is in seconds. This gives the estimate $\lambda_\infty = 4 \times
10^{-7}$~s$^{-1}$ (or $\lambda_\infty = 0.035$~day$^{-1}$). A similar
fit of Eq.~(\ref{eq:sigma_convergence}) to the standard deviation of
the Lyapunov exponents gives $\Delta = 5 \times 10^{-7}$~s$^{-1}$. We
have not attempted to quantify the uncertainties of these estimates,
but, because of the slow convergence, the uncertainties will be
relatively large. It could be expected that $P(\lambda, t)$ would
become gaussian when $\sigma_\lambda << \lambda_\infty$, or when $t >>
\Delta/\lambda_\infty^2$. For the OI flow this will occur when $t >>
40$ days, so it is clear that the integration is too short to allow
the asymptotic form of $P(\lambda, t)$ to emerge.

The mean value of the Lyapunov exponents approximated from the
stretching of an initially randomly aligned vector, $\tilde\lambda$,
is shown in Fig.~\ref{fig:lambda_mean_std} for comparison. Because of
the random alignment of the initial vector relative to the straining
of the flow, the initial stretching is zero. It grows rapidly, and then
slowly converges toward the actual value, $\langle\lambda\rangle$. A
least squares fit of Eq.~(\ref{eq:lambda_convergence}) to the data
between 15 and 80 days gives $\langle\tilde\lambda\rangle \sim 0.0428
/ t + 1.59 \times 10^{-4}/\sqrt{t} + 4.33 \times 10^{-7}$. This is
consistent with the estimate of $\lambda_\infty = 4 \times 10^{-7}
{\text{s}}^{-1}$.

\section{\label{sec:sst}Sea-surface temperature}

\begin{figure}
  \includegraphics{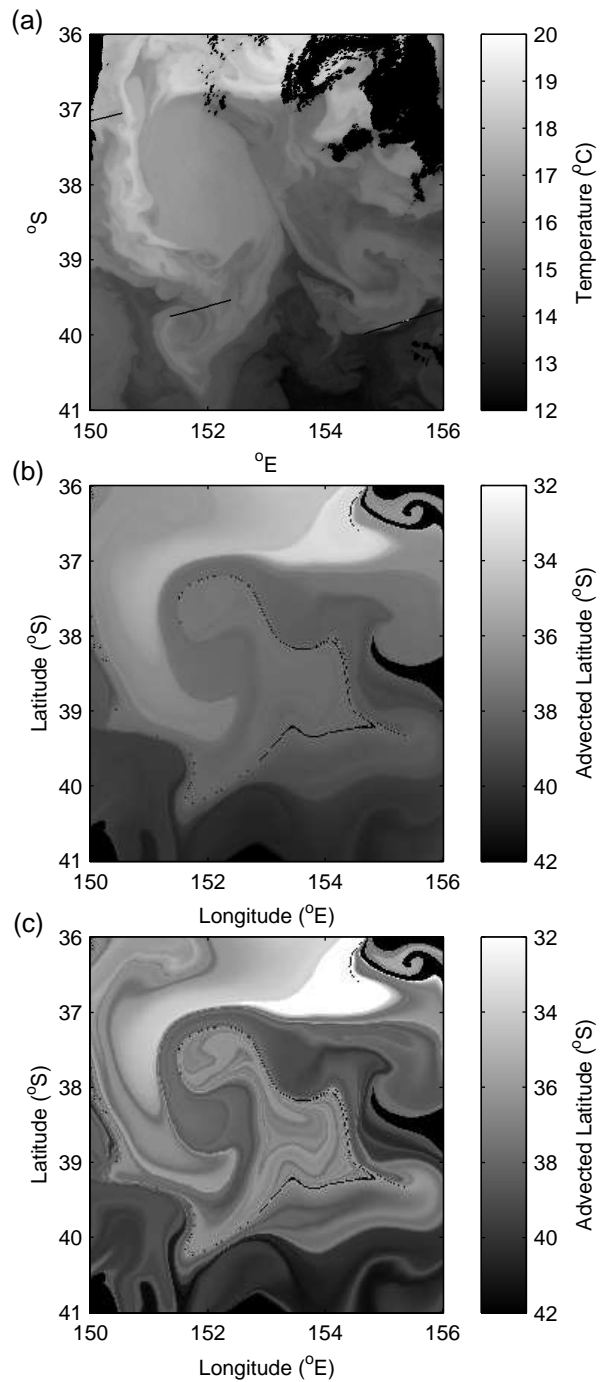}
\caption{\label{fig:sst} A satellite sea-surface temperature image
  from 22 November 1997 (a) is compared with SST modelled using the OI
  velocity field and a first-order equation for a reactive tracer,
  Eq.~(\ref{eq:dcdt}). The two modelled SST images were generated with
  the relaxation rates (b) $\alpha = 0.05$ day$^{-1}$ and (c) $\alpha
  = 0$. In (a) black marks either cloud or invalid data, and in (b)
  and (c) black marks trajectories which went outside the area in
  which the OI velocities are defined. The SST image has 1 km pixels,
  and the modelled data has a 2.2 km resolution.}
\end{figure} 

In this section the formalism developed by Neufeld {\it et al.~}
\cite{Neufeld00} and presented in section \ref{sec:scalar_spectra} is
applied to a satellite image of sea-surface temperature (SST) to test the
applicability of the theory in this context. In Fig.~\ref{fig:sst} an
SST image from 22 November 1997 is compared with images produced by
the advection of a first-order tracer by the OI flow. The OI
streamfunction for this day is shown in Fig.~\ref{fig:lyapunov_hires},
with the dominant feature being an anti-cyclonic eddy. The equation
for the tracer (Eq.~\ref{eq:dcdt}) is integrated along the
trajectories, which were calculated backwards from a regular grid with
a 2.2 km spacing. The integration is only for 40 days as for longer
times too many trajectories leave the area in which the OI velocities
are defined. The function $C_0$ is taken to be a linear gradient given
by the latitude of the trajectory. The tracer field along each
trajectory is initialised at its starting latitude. With a relaxation
time of $1/\alpha = 20$ days (Fig.~\ref{fig:sst}(b)) the broad
structures of the modelled SST compare well with what is seen in the
satellite image taken on that day (Fig.~\ref{fig:sst}(a)). The
anti-cyclonic eddy appears as an area of low gradient, with the
northern boundary being marked by a line of strong gradient at
37$^\circ$S. A tongue of warmer northern water comes down the western
side of the eddy, and there is a meandering front between latitudes
39$^\circ$S and 40$^\circ$S. Given the extreme simplicity of the
sea-surface temperature model the agreement between these images is
striking. There are, however, many smaller features which are not
resolved in the OI velocity field and so the contours of the modelled
SST are smoother than the contours of the actual SST image. If the
relaxation rate is set to $\alpha = 0$, and the latitude is advected
as a conserved tracer, then the same broad patterns are seen
(Fig.~\ref{fig:sst}(c)), but the tracer structure becomes more
striated than the observed sea-surface temperature. The highest tracer
gradients are in the regions of greatest stretching, and the filaments
of the $t = 40 $ day Lyapunov exponents
(Fig.~\ref{fig:lyapunov_hires}) are aligned with the contours of the
tracer (Fig.~\ref{fig:sst}(c)).

\begin{figure}
  \includegraphics{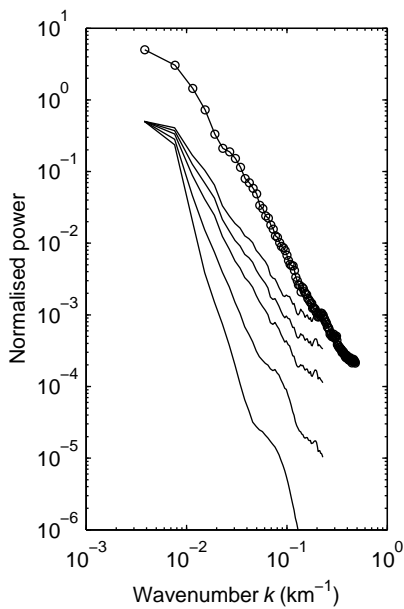}
\caption{\label{fig:power_spectra} Power spectra of the actual and
  modelled SST data shown in Fig.~\ref{fig:sst}. The spectra are the
  mean of the 1-D power spectra taken along 250 km long sections of
  constant latititude which contain no invalid data. The normalisation
  is arbitray, having been chosen to aid comparison of the spectra.
  The spectrum of the actual SST data is shown by the line with
  circles. The other lines correspond to the modelled SST, with the
  following values of $\alpha$ (day$^{-1}$), from the steepest to the
  flattest: 0.2, 0.1, 0.05, 0.025 and 0.}
\end{figure}

The spatial patterns in these images may be quantitatively compared
through their power-spectra, Fig.~\ref{fig:power_spectra}. The spectra
were taken from the 250 km long constant-latitude sections through the
image which only contained valid data. The spectra shown are the
averaged cyclic power-spectra, with the sections being detrended and
multiplied by a Hann filter before calculating the periodogram. The
averaged power-spectra have been given an arbitrary normalisation to
allow them to be readily compared. Between inverse wavenumbers of 10
and 100 km the power spectra have a power-law form, with there being a
progressive steepening of the spectra as $\alpha$ increases. This has
previously been seen in simulations of a first-order tracer in a model
flow intended to represent mesoscale turbulence \cite{Abraham98}, and
is as expected from theory (Eq.~\ref{eq:beta_monofractal}). The
power-law exponents of the spectra in Fig.~\ref{fig:power_spectra} are
$\beta = 2.44 \pm 0.04$ (actual SST); $\beta = 2.12 \pm 0.03$
(conserved tracer); $\beta = 2.28 \pm 0.03$ ($1/\alpha$ = 40 days);
$\beta = 2.46 \pm 0.04$ ($1/\alpha$ = 20 days); $\beta = 2.77 \pm
0.09$ ($1/\alpha$ = 10 days); and $\beta = 3.47 \pm 0.14$ ($1/\alpha$
= 5 days). The spectral exponent of the actual SST data is not
significantly different from the exponent of the modelled data with
$1/\alpha = 20$ days, suggesting that the SST is responding to changes
in forcing with a timescale of around 20 days.

The spectrum of the conserved tracer is flatter than the tracer with
$1/\alpha = 40$ days but, because of the short integration time, the
spectrum is still steeper than the $k^{-1}$ form expected for a forced
conserved tracer. The 40 day period over which the trajectories are
defined is too short to allow a full transfer of variance from low to
high wavenumbers. The timescale expected for the transfer between two
lengthscales $L_{\max}$ and $L_{\min}$ is $T_L =
\log(L_{\max}/L_{\min})/\lambda_\infty$. For the OI flow
$1/\lambda_\infty$ = 29 days, so with $L_{\max}/L_{\min} = 10$ the
timescale for variance transfer through the mesoscale is $T_L = 70$
days. An integration of at least this length would be needed to allow
a spectrum with a slope close to $k^{-1}$ to be established. The 40
day trajectory length is too short to allow the relation between
spectral-exponent and the relaxation-rate, given in
Eq.~(\ref{eq:beta_monofractal}), to be tested.
\begin{figure}
  \includegraphics{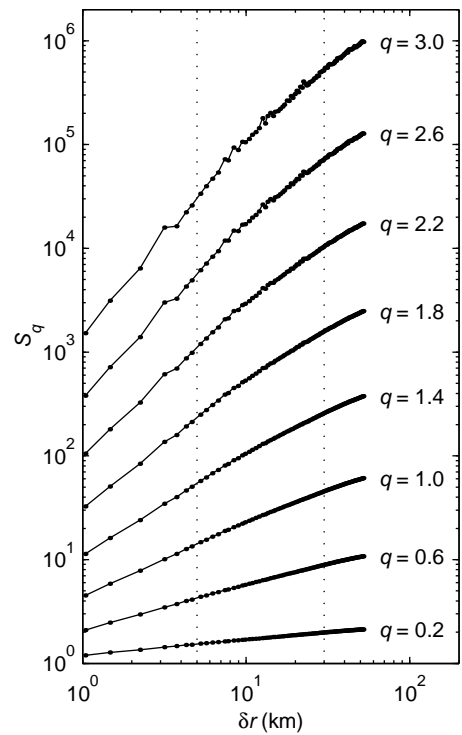}
\caption{\label{fig:structure} Structure functions, $S_q(\delta r)$,
  calculated using Eq.~(\ref{eq:Sq}) from the November 22 SST image
  (Fig.~\ref{fig:sst}(a)). The scaling exponents, $\zeta(q)$, are
  calculated from a least-squares fit to the structure functions
  within the range 5 to 30 km, shown by vertical dotted lines. The
  structure functions are approximately power-law over this range, but
  roll off toward larger separations, $\delta r$.}
\end{figure}

\begin{figure}
  \includegraphics{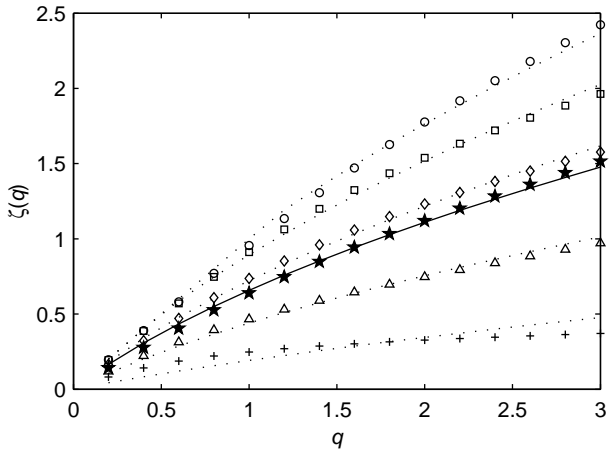}
\caption{\label{fig:zeta} Multifractal scaling exponents of
  sea-surface temperature. The solid stars ($\star$) are the exponents
  calculated from the SST data shown in Fig.~\ref{fig:sst}(a). The
  solid line is a least-squares fit of Eq.~(\ref{eq:zetaq_ld}) to the
  exponents. The other symbols and the dotted lines mark the scaling
  exponents calculated from the modelled data and the associated
  best-fit curve, with the following values of $\alpha$ (day$^{-1}$):
  0.2 (circle), 0.1 (square), 0.05 (diamond), 0.025 (triangle), 0 (plus).}
\end{figure}

The multifractal scaling of a tracer may be calculated using the
equation for the structure function (Eq.~\ref{eq:Sq}). The structure
functions of the SST image shown in Fig.~\ref{fig:sst}(a) and of
simulated SST were calculated for $\delta r < 50$ km, and with $q$
between 0.2 and 3 (Fig. \ref{fig:structure}). The modelled data were
calculated as for Fig.~\ref{fig:sst}(b, c), but with an 80 day
integration time and with a range of relaxation rates $\alpha$. The
longer trajectory lengths meant that 25\% of the trajectories went
outside the area where the velocities were defined. Despite the
presence of the invalid data, it was still possible to calculate the
structure functions as the averaging is only taken over pairs of valid
points. The data were detrended before the calculation was made.
The resulting scaling exponents, obtained from a least squares
regression of $\log(S_q(\delta r))$ against $\log(\delta r)$, are
shown in Fig.~\ref{fig:zeta}.  The structure functions of the SST
image lie close to the curve derived from the simulated SST with
$1/\alpha = 20$ days, in agreement with the estimate of $\alpha$
obtained by comparing power-spectra. A least-squares fit of the
relation given by Eq.~(\ref{eq:zetaq_ld}) closely follows the SST
data, with the best fit dimensionless parameters being
$\alpha/\lambda_\infty = 0.90$ and $\Delta/\lambda_\infty = 1.12$.
With the value estimated for the OI flow of $\lambda_\infty = 4 \times
10^{-7}$~s$^{-1}$ this results in $\alpha = 0.031$~day$^{-1}$, or
$1/\alpha = 32$~days, and $\Delta = 4.5 \times 10^{-7}$~s$^{-1}$.
There is a long chain of assumptions required to allow this estimate
to be made, but it doesn't require an explicit modelling of the flow
or, indeed, any knowledge of the flow beyond $\lambda_\infty$.
Although it is perhaps coincidental, it is pleasing that the value of
$\Delta$ is close to that found from an analysis of the
finite-time Lyapunov exponents. This suggests that the spatial
structure seen in the satellite image is consistent with the
representation of SST as an advected first-order tracer. The other dotted lines in
Fig.~\ref{fig:zeta} result from fitting the expression for $\zeta(q)$
to estimates made from the simulated SST. The values of $1/\alpha$
(days) predicted from the best-fit curves are 21 (5); 24 (10); 29 (20)
and 50 (40), where the actual value is shown in brackets. The curve
fitting over-estimates the relaxation time-scale, particularly when
the relaxation is rapid. This is likely to be because the derivation
of the expression for the fitted curve (Eq.~\ref{eq:zetaq_ld}) assumes
that $P(\lambda, t)$ is gaussian. This is a poor approximation at
small times.

Is a 20 day response time for sea-surface temperature reasonable? A
one-dimensional turbulence closure model was recently used to simulate
seasonal variability of SST in the eastern Tasman Sea, at a similar
latitude to the data analysed here, but close to New Zealand
\cite{Hadfield00}. The model used heat and wind forcing derived from
meteorological observations. It was found that a perturbation of the
atmospheric forcing, applied so that the perturbation was removed at
the end of November, had an effect which decayed with a time-scale of
one to two months. The response was more complex than a simple first
order relation, but nevertheless the response-time inferred from the
water column model was of a similar order to that found here. It would
be interesting to combine the Lagrangian advection with a water-column
model that included realistic heat and wind forcing.

A similar set of assumptions about the dynamics of sea-surface
temperature has been used in mesoscale simulations of SST variability
by Klein and Hua \cite{Klein90}. The focus of these simulations was on
the response of SST to episodic bursts of wind. When the mixed-layer
is deepened by a wind event the surface temperature is imprinted with
a signal from the sub-surface ocean. The mesoscale dynamics below the
thermocline may be represented by quasi-geostrophic turbulence. In the
ocean interior, temperature is not passive and is dynamically
constrained, with a predicted spectral exponent of $\beta=3$ in the
sub-mesoscale range. So, when the mixed-layer deepens the sea-surface
temperature spectra steepen towards this value.  During subsequent
stirring of the mixed-layer it was found that temperature variance was
transferred towards higher wave-number, resulting in the spectral
exponent becoming smaller with time. From this point of view, the
processes determing sea-surface temperature spectra are a steepening
through episodic wind-mixing and a flattening through horizontal
stirring.  Following this argument, sea-surface temperature spectra
with exponents less than one should be possible during periods of time
when the mixed-layer is not deepening for many months, such as over
summer. Such flat spectra are not usually observed. An analysis of the
spatial pattern in a time-series of SST imagery would allow the
relative importance of continuous relaxation and episodic wind-mixing
to be resolved.

\section{\label{sec:discussion}Discussion}

Estimates or descriptions of stirring in the surface ocean are rare.
In this paper, a spatially and temporally complete velocity dataset
was used to calculate the Lyapunov exponents of the flow in a region
of the East-Australian current. The mean stirring rate was found to be
$\lambda_\infty = 4 \times 10^{-7}$ s$^{-1}$, corresponding to a
timescale of $1/\lambda_\infty = 29$ days. The velocity field used is
derived using relatively indirect techniques, and only resolves the
largest mesoscale features. The question then arises, how reliable is
the estimate of the stirring?  The observation that tracers advected
in ocean flows are filamental suggests that there is a separation
between the diffusive length scale and the scale which is controlling
the stirring. If, indeed, the larger mesoscale features are dominating
the shear then the analysis will be appropriate. This will not be the
case in shallow waters, where the eddy length scale is too small to be
resolved using the remote methods that are relied on here.  There have
been beautiful observations made of sub-mesoscale stirring, seen in
photographs taken from space of sun-glitter on the sea-surface
\cite{Munk00, Munk01}. These show small eddies, with diameters of only
tens of kilometers, organising surface flotsam into filamental slicks.
It appears from these photographs that sub-mesoscale eddies may at
times dominate the stirring of tracers. In this case, the methods used
here would underestimate stirring rates.

From the distribution of the Lyapunov exponents it is clear there is a
very wide range in the stretching rates which could be experienced by
a patch of tracer in the ocean. Moreover, it would be difficult to
predict the stretching of a patch from an Eulerian snapshot of the
flow. While the analysis used here is based on a velocity field which
could, in principle, be obtained in near real time, predicting the
dispersal of either a pollutant or a deliberately released tracer is
unlikely to be possible. Small errors in the velocity will lead to a
rapid divergence of trajectories. Despite the spread of the
probability distribution of Lyapunov exponents it is intriguing to
note that the only previous estimate of stirring in the surface ocean,
obtained from a satellite image of a deliberately induced
phytoplankton bloom \cite{Abraham00}, found a value for the stretching
of $\lambda(42\quad \hbox{days}) = 6 \times 10^{-7}$ s$^{-1}$. Other
estimates of mesoscale stirring from a numerical model ($5.8 \times
10^{-7}$ s$^{-1}$ \cite{Haidvogel84}) and from a deep tracer release
($3 \pm 0.5 \times 10^{-7}$ s$^{-1}$\cite{Ledwell93, Ledwell98}) are
all of a similar order. The agreement of the values obtained, using
very different methods, suggests that the estimate for
$\lambda_\infty$ found by analysing the OI velocity field is a
reasonable one, with the advantage over the tracer-releases that the required data are
readily available.

There has been much work recently on the stirring of reactive tracers,
particularly phytoplankton, in the surface ocean \cite{Franks97,
  Abraham98, Bracco00, Martin00, Scheuring00, Karolyi00, Lopez01,
  Richards01, McLeod01}.  This research has generally relied on simple models
of the ocean flow. The study of Lagrangian chaotic flows, in
particular, has allowed the problem to be broken into its simplest
components and understanding of the patterns seen in advected tracers
has been advanced considerably.  The application of this theoretical
work has been somewhat hampered by the lack of appropriate oceanic
data. The analysis presented here is intended to help alleviate that
problem, but is presented as a simple case study only. There are many
further directions in which it may be taken.  It would be interesting
to test the application of the theoretical relationships in a model of
ocean turbulence which was not disadvantaged by the short integration
times that we were restricted to here. As far as sea-surface
temperature is concerned, it would be interesting to carry out a
systematic analysis of the whole East-Australian
optimally-interpolated dataset. There may be seasonal variations in
the estimated value of the response time, $\alpha$, and if these
reflect real ocean processes then they will contain information on
mixed-layer depth. Since mixed-layer depth is a crucial factor
controlling primary production in the ocean, a method for deriving it
 from remotely sensed data would have great value.  As more satellite
data becomes available, and as techniques for extracting sea-surface
velocities from the temperature data are improved, it is hoped that
well resolved surface velocity fields will become widely avaible.
Analysis of these will transform our understanding of dispersal
processes in the surface ocean.

\begin{acknowledgments}
  
  E.A. is grateful to the organisers of the Los Alamos workshop on
  Active Chaotic Flow (2001) for their invitation to attend. The work
  presented here was motivated by material presented at the workshop.
  The research was funded by the New Zealand Foundation for Research,
  Science and Technology, as part of the Ocean Ecosystems programme.

\end{acknowledgments}

\bibliography{chaotic_stirring}

\end{document}